\documentclass[twocolumn,superscriptaddress]{revtex4-2}
\usepackage[T1]{fontenc}
\usepackage{times}
\usepackage[utf8]{inputenc}
\usepackage[resetlabels,labeled]{multibib}
\usepackage{enumitem}
\usepackage{amsmath}
\usepackage{amsthm}
\usepackage{amsfonts}
\usepackage{amssymb}
\usepackage{bm}
\usepackage{bbm}
\usepackage{bbold}
\usepackage{graphicx,epsfig,color}
\usepackage{physics}
\usepackage{floatrow}
\usepackage{float} 
\usepackage{mathtools}
\usepackage{mathrsfs}
\usepackage{latexsym}
\usepackage{color}
\definecolor{color1}{rgb}{0,0,0.7}
\usepackage[colorlinks=true,linkcolor=teal,citecolor=purple,urlcolor=color1]{hyperref}
\usepackage{cleveref}
\usepackage{extarrows}
\usepackage{tikz}
\usepackage[ruled,vlined]{algorithm2e}
\usepackage{bm}

\newcommand{\ot}{\otimes}
\newcommand{\ms}[1]{\mathsf{#1}}

\definecolor{plbred}{rgb}{0.8,0,0}
\definecolor{mplcol}{rgb}{0.,0.8,0.}

\newcommand{\On}[1]{\mathcal{O}\left(\frac{1}{n^{#1}}\right)}
\begin{document}

\title{Minimizing Dissipation via Interacting Environments: Quadratic Convergence to Landauer Bound}
\author{Patryk Lipka-Bartosik}
 \affiliation{Department of Applied Physics, University of Geneva, 1211 Geneva, Switzerland}
\author{Mart\'i Perarnau-Llobet}
\affiliation{F\'isica Te\`orica: Informaci\'o i Fen\`omens Qu\`antics, Department de F\'isica, Universitat Aut\`onoma de Barcelona, 08193 Bellaterra (Barcelona), Spain}
 \affiliation{Department of Applied Physics, University of Geneva, 1211 Geneva, Switzerland}

\date{\today}

\begin{abstract}
We explore the fundamental limits on thermodynamic irreversibility when cooling a quantum system in the presence of a finite-size reservoir. First, we prove that for any non-interacting $n$-particle reservoir, the entropy production $\Sigma$ decays at most linearly with $n$. Instead, we derive a cooling protocol 
in which  $\Sigma \propto 1/n^2$, which is in fact the best possible scaling. This becomes possible due to the presence of  interactions in the finite-size reservoir, which must be prepared at the verge of a phase transition. Our results  open the possibility of cooling  with a higher energetic efficiency via interacting reservoirs. 
\end{abstract}

\keywords{}
\maketitle

\section{Introduction}



Thermodynamics was originally born as a theory to characterize macroscopic heat engines. Since then it has vastly increased its regime of applicability, nowadays contributing to our understanding of the physics of systems ranging from black holes to biological systems. In particular, the thermodynamics of quantum systems is being intensively investigated in the growing fields of stochastic and quantum thermodynamics~\cite{Esposito2009,Jarzynski2011,Seifert2012,Goold2016,Vinjanampathy2016,Mitchison2019,myers2022quantum}.   
In such systems, the thermal reservoir is typically described via a Markovian master equation satisfying detailed balance~\cite{breuer2002theory,rivas2012open}, 
which enables a simple and efficient description of its effect. 
Going beyond this picture is however important in quantum systems, where strong coupling, non-Markovian as well as finite-size effects are often present. In order to understand, but also to exploit, such effects in thermodynamics, a more complete description of the system-reservoir dynamics is needed~\cite{Trushechkin2022,Ptaszyifmmode2019Entropy,Gour2015,czartowski2023thermal}.

At a fundamental level, this microscopic description  enables
the understanding of the fundamental limits of thermodynamic processes~\cite{janzing2000thermodynamiccostreliabilitylow,Horodecki_2013,Brandao2013,biswas2022fluctuation,PhysRevX.8.041051,PhysRevX.11.011061,junior2022geometric,bera2017generalized,bera2019thermodynamics}, including ultimate bounds to  cooling~\cite{Masanes2017,Alhambra2019,Wilming_2017,PRXQuantum.4.010332,lipka2023operational} and finite-size corrections to the Landauer's principle~\cite{reeb2014improved,Timpanaro2020Landauer}. 
At a more practical level,  
this  approach enables exploration of more exotic or engineered reservoirs for stochastic and quantum thermodynamics.   Relevant examples   are finite size reservoirs~\cite{reeb2014improved,Richens2018finite,Moreira2023Stochastic,Schaller2014Relaxation,Yuan2022Optimizing,Amato2020Noninteracting,Ma2020Effect}, which can be exploited for ultraprecise quantum calorimetry~\cite{Gasparinetti2015Fast,Karimi2020}, as well as strongly coupled  \cite{Trushechkin2022,Newman2017Performance,Nazir2018,Strasberg2019Repeated,Pancotti2020,Carrega2022Engineering}, non-Markovian~\cite{Ptaszyifmmode2022}, superradiant~\cite{Hardal2015,Niedenzu2018,Manzano2019,Kloc2019Collective,Watanabe2020Quantum,Kloc2021,Tajima2021Superconducting}, 
and non-equilibrium reservoirs~\cite{Scully2003,Rossnagel2014,Manzano2016Entropy,Manzano2018Quantum}. 

\begin{figure}[h!]
    \centering
    \includegraphics[width=0.9\linewidth]{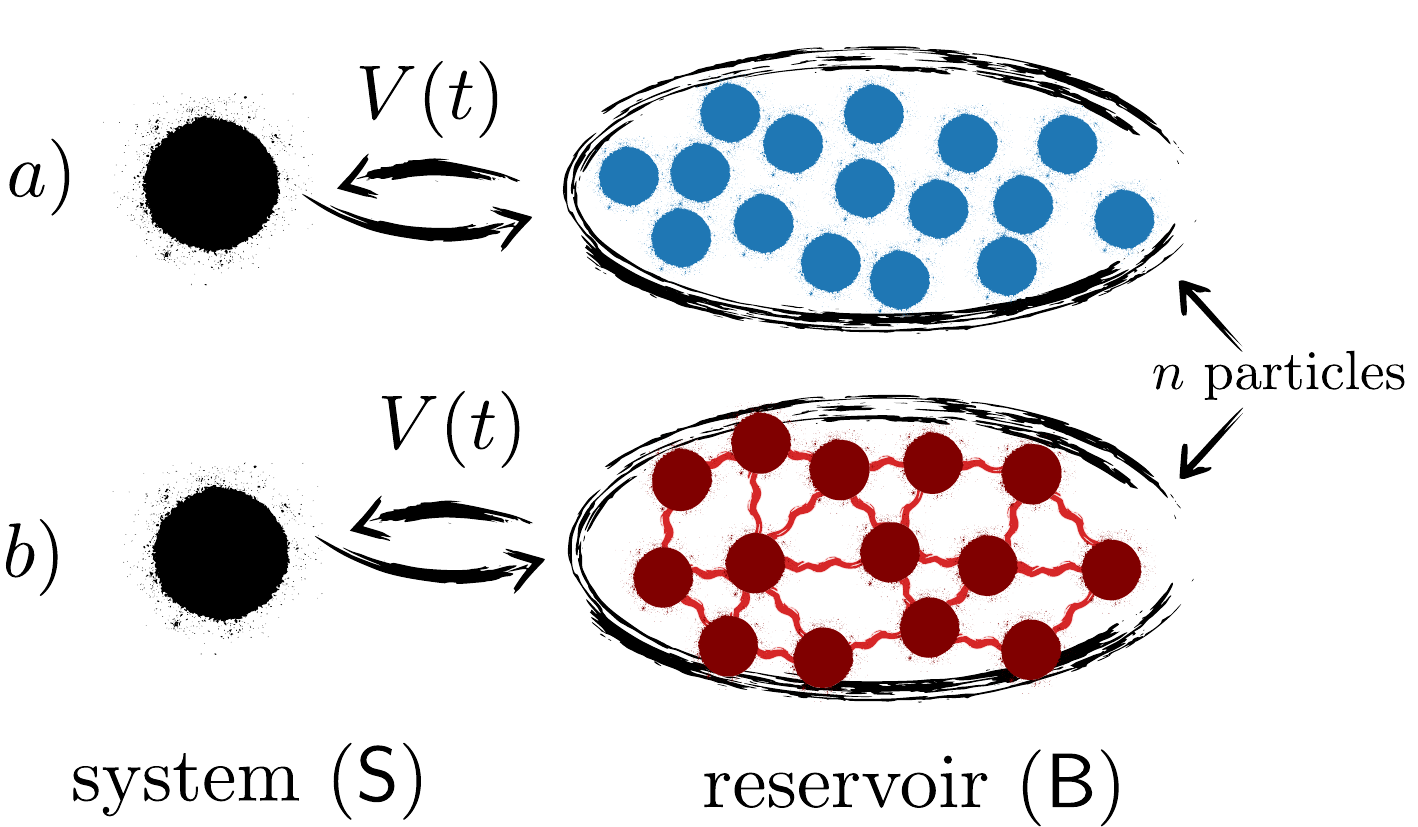}
    \caption{\textbf{Framework.} a) A system (S) is cooled by external control $V(t)$ combined with the interaction with a finite-size reservoir (B) made-up of non-interacting components. b) In this case, the thermal reservoir features strong interactions between its components. We prove here that this can be exploited for more efficient cooling of the system.  }
    \label{fig:enter-label}
\end{figure}

\begin{figure}
    \centering
    \includegraphics[width=\linewidth]{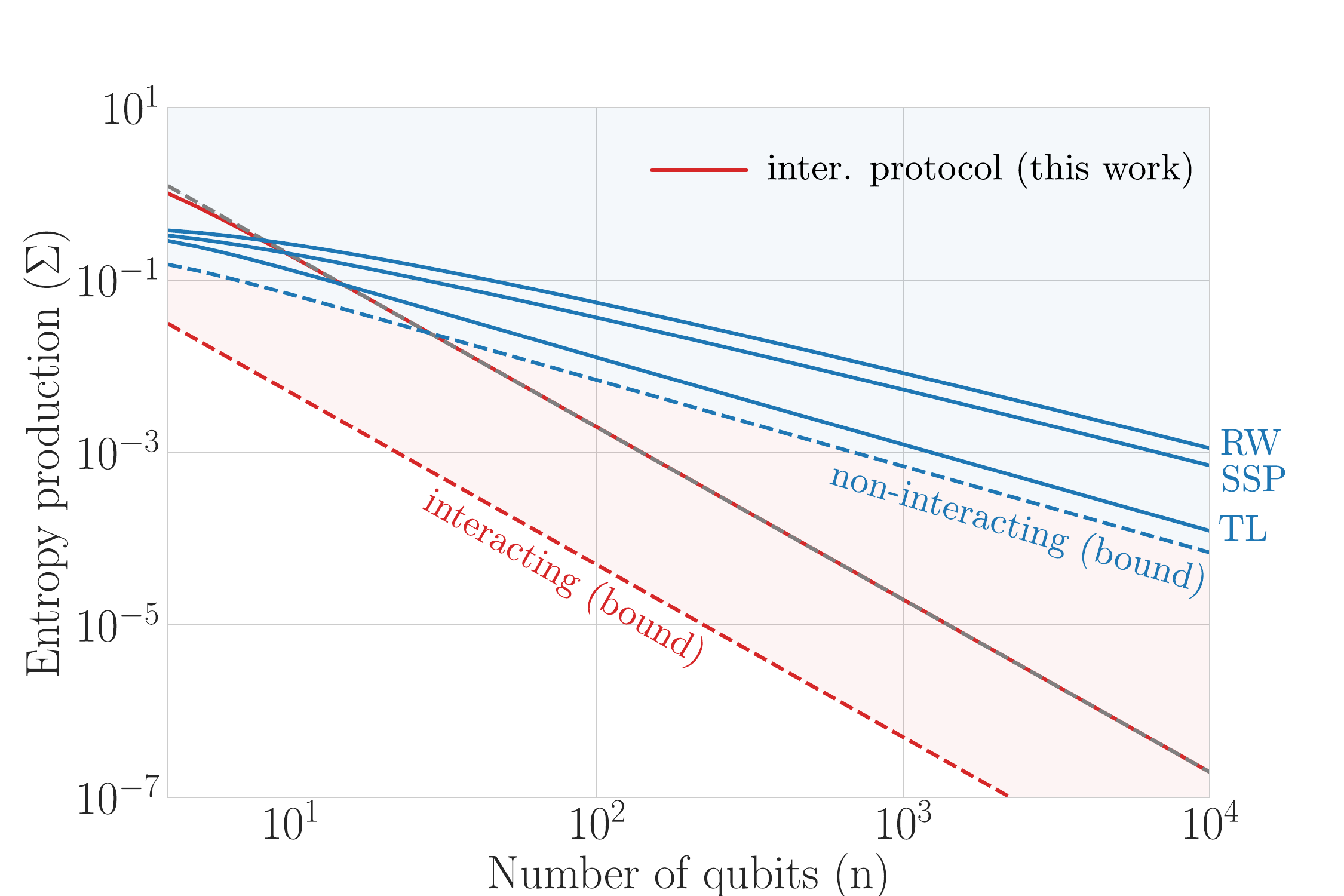}
    \caption{\textbf{Thermodynamic processes for Landauer erasure.} The solid lines correspond to entropy production in thermodynamic processes. RW, SSP and TL correspond to collisional processes described, respectively, in Refs. \cite{reeb2014improved}, \cite{skrzypczyk2014work} and \cite{taranto2024efficiently}. Red curve corresponds to the process using an interacting environment described in Sec. \ref{sec:non-inter}. Dashed lines correspond to analytic bounds, i.e. lower-bound for non-interacting environments \eqref{eq:noninter_bound} (blue), lower-bound for general environments \eqref{eq:rw_bound} (red) and an upper-bound for the thermodynamic process discussed in Sec. \ref{sec:non-inter} [see Eq. \eqref{eq:diss_protocol}] (grey). Parameters used: $\beta = 1$, $q = n^{-\alpha}$ with $\alpha = 3$.}
    \label{fig:1}
\end{figure}

 The main goal of this paper is to understand the potential of  interacting reservoirs to reduce the energetic cost of thermodynamic processes, particularly cooling. 
We consider a reservoir made up of $n$ particles which can feature strong interactions among themselves, prepared in a Gibbs state at inverse temperature $\beta$ (see Fig.~\ref{fig:1}). We then couple it to a quantum system and apply a general unitary operation to both systems, aiming to decrease the entropy of the quantum system.
The second law of thermodynamics bounds the entropy change $\Delta S$ of the system and the energy dissipated into the reservoir $Q$, namely
\begin{equation} \label{eq:entr_prod}
    \Sigma := \beta Q + \Delta S \geq 0,
\end{equation}
where $\Sigma$ is the so called entropy production~\cite{Landi2021Irreversible}, and can reach zero only in the limit $n\rightarrow \infty$~\cite{reeb2014improved}. 
Given the task of cooling a single qubit, several works have analyzed the decay of $\Sigma$ with $n$~\cite{reeb2014improved,skrzypczyk2014work,Baumer2019imperfect}, 
finding a convergence  of the form 
$\Sigma  = A/n$ for $n\gg 1,$ where $A$ depends on the specific protocol (see Fig. \ref{fig:1}). 
A recent work~\cite{taranto2024efficiently} minimized the constant for a class of collisional models, finding $A \approx \pi^2/8$ using tools from thermodynamic geometry~\cite{Nulton1985,Abiuso2020Geometric}. Here we prove that indeed  $\Sigma$ can decay at most as $\mathcal{O}(1/n)$ for all non-interacting reservoirs, finding a general lower bound, Eq.~\eqref{eq:noninter_bound}, which for a single qubit reads
\begin{align}   
   \Sigma  \geq  \frac{1}{3n},  \hspace{10mm} \rm{(non-interacting)}.
\end{align}
Instead, we show that interacting reservoirs surpass this bound, and construct an explicit reservoir Hamiltonian and cooling process  achieving
 \begin{align} 
    \Sigma =  \frac{2 \pi^2}{n^2} \qquad \text{for} \quad n\gg 1. 
\end{align}
We further argue that this is, in fact, the best possible scaling with $n$ for any thermodynamic process.
 This provides a rigorous example in which an interacting engineered reservoir outperforms \emph{any} non-interacting one of the same size (see Fig. \ref{fig:1}). Our results hence open the possibility of cooling, or information erasure, with a higher energetic efficiency by exploiting interacting nanoscale reservoirs.  

\section{Framework}
A thermodynamic system $\ms{S}$ is characterized by its Hamiltonian $H_{\ms{S}}$ and density matrix $\rho_{\ms{S}}$. We say that a system is in thermal equilibrium at (inverse) temperature $\beta$ if its density matrix can be written as $\tau_{\ms{S}} := e^{-\beta H_{\ms{S}}} / Z_{\ms{S}}$, with $Z_{\ms{S}} := \Tr[e^{-\beta H_{\ms{S}}}]$ being the partition function. A \emph{thermodynamic process} $\mathcal{P} = (\rho_{\ms{S}}, U, H_{\ms{B}})$ is a process in which system $\ms{S}$ interacts with an environment $\ms{B}$ prepared in a Gibbs state $\tau_{\ms{B}}$ via a unitary mechanism described by a time propagator $U$. This process is modelled using a Hamiltonian $H_{\ms{SB}}(t)=H_{\ms{S}}+H_{\ms{B}} + V(t)$, where $H_{\ms{S}/\ms{B}}$ are local system/environment Hamiltonians and $V(t)$ is a cyclic potential such that $U\,=\, \mathcal{T}[\text{exp}(-i\int_{0}^{\tau} H_{\ms{SB}}(t) \mathrm{d}t )]$ with $\mathcal{T}$ being the time-ordering operator. The two systems are assumed to be initially uncorrelated, i.e. $\rho_{\ms{SB}} = \rho_{\ms{S}} \ot \tau_{\ms{B}}$. 
During a thermodynamic process the joint system $\ms{SB}$ is transformed as
\begin{align} \label{eq:protocol}
    \rho_{\ms{S}} \ot \tau_{\ms{B}} \rightarrow \sigma_{\ms{SB}} := U(\rho_{\ms{S}} \ot \tau_{\ms{B}})U^{\dagger}.
\end{align}
The von Neuman entropy is defined for a density matrix $\rho$ as $S(\rho) := - \Tr \rho \log \rho$. In this way the change of entropy of $\ms{S}$ under a thermodynamic process is given by $\Delta S_{\ms{S}} := S(\sigma_{\ms{S}}) - S(\rho_{\ms{S}})$, where $\sigma_{\ms{S}} := \Tr_{\ms{B}}\sigma_{\ms{SB}}$ denotes the partial trace.

In any thermodynamic process the heat $Q := \Tr[H_{\ms{B}}(\sigma_{\ms{B}} - \tau_{\ms{B}})]$ dissipated to the environment is bounded by $\beta Q + \Delta S_{\ms S} \geq  0$. 
This inequality can be sharpened to the following equality~\cite{Esposito2010finite,reeb2014improved}
\begin{align} \label{eq:diss}
   \beta Q + \Delta S_{\ms S} = I(\ms{S}:\ms{B})_{\sigma} + D(\sigma_{\ms{B}}\|\tau_{\ms{B}}) =:  \Sigma,
\end{align}
where $D(\rho\|\sigma) := \Tr[\rho(\log \rho - \log \sigma)]$ is the (quantum) relative entropy \cite{umegaki1962conditional} and $I(\ms{S}:\ms{B})_{\sigma} := D(\sigma_{\ms{SB}}\| \sigma_{\ms{S}} \ot \sigma_{\ms{B}})$ is the (quantum) mutual information. 
The entropy production $\Sigma$ quantifies the thermodynamic irreversibility of the process~\cite{Landi2021Irreversible}. In what follows we will be interested in minimizing this quantity in the case when the dimension of the environment, $d_{\ms{B}}$, is finite. This will allow us to study the corrections to the second law (with focus on the Landauer bound) resulting from the nature of the environment used. 


\section{Results}

In this work we explore the limits on the minimisation of $\Sigma$ as a function of the reservoir's size, which is composed of $n$ possibly interacting qubits. In Table \ref{tab:1}, we summarize our results for the case of Landauer erasure and put them in context with the state of the art.   

\begin{table}[]
    \centering
    \begin{tabular}{|c|c|c|c|}
         \hline
           & Lower bound & Best known protocol  \\
         \hline
         Non-interacting & $\frac{1}{3} n^{-1}$ \hspace{5pt} (\textbf{this work}) & $\frac{1}{8}\pi^2 n^{-1}$ \hspace{5pt} (Ref. \cite{taranto2024efficiently})   \\
         \hline
         Interacting &$\frac{1}{\log 2} n^{-2}$ \hspace{5pt} (Ref. \cite{reeb2014improved}) & $2 \pi^2 n^{-2}$ \hspace{5pt} (\textbf{this work}) \\
         \hline
    \end{tabular}
    \caption{\textbf{Entropy production during Landauer erasure in finite-size environments.} The table shows the lower bounds and best known processes for the particular case of the erasure (or cooling) of a single qubit with a $n$-qubit environment. The lower bounds are known for arbitrary systems and reservoirs, leading to the same scaling with $n$ but with a different constant (see Ref.~\cite{reeb2014improved} and text below). 
    } 
    \label{tab:1}
\end{table}

\subsection{Entropy production in non-interacting thermal environments} 
Consider a thermal environment composed of $n$ subsystems, each with local dimension $d$ and described by local Hamiltonian $h_{\ms{B}}^{(i)}$. In the absence of interactions the total Hamiltonian of the environment reads 
\begin{align}\label{eq:h_noninter}
    H_{\ms{B}}^{\text{no-int}} = \sum_{i=1}^n h_{\ms{B}}^{(i)}.
\end{align}
Now consider a thermodynamic process of the form \eqref{eq:protocol} with $H_{\ms{B}} = H_{\ms{B}}^{\text{no-int}}$. The entropy production $\Sigma$ in any such process in given by Eq. \eqref{eq:diss}. We will now argue that the form of the Hamiltonian in Eq. \eqref{eq:h_noninter} implies a lower bound on entropy production in any thermodynamic process (for any $\rho_{\ms{S}}$ and $U$).   

For that we define a density operator $\omega_{\ms{B}}$ such that $\omega_{\ms{B}} \propto e^{-\beta^{\star} H_{\ms{B}}}$. The parameter $\beta^{\star}$ is chosen so that the new state has the same average energy as the state of the environment $\sigma_{\ms{B}}$ at the end of the process, that is $\tr[H_{\ms{B}}\omega_{\ms{B}}] = \tr[H_{\ms{B}}\sigma_{\ms{B}}]$. Now observe that the relative entropy satisfies the inequality \cite{csiszar1975}:
\begin{align}
\label{eq:nonint_firststep}
    D(\sigma_{\ms{B}}\| \tau_{\ms{B}}) =  D(\sigma_{\ms{B}}\| \omega_{\ms{B}}) +  D(\omega_{\ms{B}}\| \tau_{\ms{B}}) \geq  D(\omega_{\ms{B}}\| \tau_{\ms{B}}),
\end{align}
which, given that both $I(\ms{S}:\ms{B})_{\sigma}$ and $D(\sigma_{\ms{B}}\|\omega_{\ms{B}})$ are non-negative, leads to the lower bound $\Sigma \geq D(\omega_{\ms{B}}\| \tau_{\ms{B}})$. Using the fact that $\ms{B}$ is composed of $n$ non-interacting particles [see Eq. \eqref{eq:h_noninter}] we can further write 
\begin{align} \label{eq:diss1}
    \Sigma \geq \sum_{i=1}^n D(\omega_{\ms{B}}^{(i)}\| \tau_{\ms{B}}^{(i)}).
\end{align}
where $\tau_{\ms{B}}^{(i)} = e^{-\beta h_{\ms{B}}^{(i)}}/\tr(e^{-\beta h_{\ms{B}}^{(i)}})$. In Appendix \ref{app1} we 
further show that the sum can be lower-bounded as
\begin{align}
    \sum_{i=1}^n D(\omega_{\ms{B}}^{(i)}\| \tau_{\ms{B}}^{(i)}) \geq \left(\frac{\Delta S_{\ms{B}}^{\star}}{\log d}\right)^2 \frac{1}{3n},
\end{align}
where $\Delta S_{\ms{B}}^{\star} := S(\omega_{\ms{B}}) - S(\tau_{\ms{B}})$. Since the Gibbs state is, by definition, a state with maximal von Neumann entropy given fixed average energy, we have $S(\omega_{\ms{B}}) \geq S(\sigma_{\ms{B}})$ and hence $\Delta S_{\ms{B}}^{\star} \geq \Delta S_{\ms{B}} := S(\sigma_{\ms{B}}) - S(\tau_{\ms{B}})$. Due to the additivity of the von Neuman entropy and unitarity of the process \eqref{eq:protocol} we also have that $\Delta S_{\ms{S}} + \Delta S_{\ms{B}} \geq 0$, which further implies that $\Delta S_{\ms{B}}^{\star} \geq - \Delta S_{\ms{S}}.$

We conclude that any thermodynamic process that changes the entropy of the system by $\Delta S_{\ms{S}}$ and uses an environment with a non-interacting Hamiltonian \eqref{eq:h_noninter} satisfies
\begin{align} \label{eq:noninter_bound}
    \Sigma \geq \left(\frac{\Delta S_{\ms{S}}^{}}{\log d}\right)^2 \frac{1}{3 n}.
\end{align}
The above bound is the first main result of this paper, demonstrating that entropy production decays at most as  $\propto 1/n$ in the presence of a non-interacting environment. 

Several thermodynamic processes that approach Landauer's bound and achieve the linear decay of $\Sigma$ have been investigated in the literature. The most well-known examples are (memory-less) collisonal processes. In such processes the environment is composed of $n$ systems of dimension $d_{\ms{S}}$, i.e. $\ms{B} = \ms{B}_1 \ms{B}_2 \ldots \ms{B}_n$ so that $d_{\ms{B}} = d_{\ms{S}}^n$.
The unitary $U$ is chosen to be a collection of successive two-body interactions between the system $\ms{S}$ and the subsequent environmental subsystems $\ms{B}_i$, that is $U = U_{\ms{S}\ms{B}_1} U_{\ms{S}\ms{B}_2} \ldots U_{\ms{S}\ms{B}_n}$. In Ref. \cite{reeb2014improved} a collisional process is proposed which achieves $\Sigma \leq A/n$ with $A:= D(\sigma_{\ms{S}}\|\rho_{\ms{S}}) + D(\rho_{\ms{S}}\|\sigma_{\ms{S}})$. Ref. \cite{skrzypczyk2014work} proposed a collisional process that achieves a slightly better constant $A$. Finally, to the best of our knowledge, the process that achieves the best constant $A$ was recently proposed in Ref. \cite{taranto2024efficiently}. There the framework of thermodynamic geometry is used to determine the optimal energy structure of the environment for minimizing entropy production.  These results are shown in Fig.~\ref{fig:1} together with our lower bound \eqref{eq:noninter_bound}, see also Table \ref{tab:1}.

\subsection{Entropy production in interacting thermal environments} \label{sec:non-inter}
Let us now shift our attention to thermodynamic processes that involve thermal environments with general, potentially interacting, Hamiltonians. In this case one can expect that the bound from Eq. \eqref{eq:diss1} can be violated and therefore lead to an advantage over non-interacting environments. As we are interested in fundamental bounds, we consider the possibility to engineer arbitrary interactions between the $n$ constituents of the environment.  Equivalently, we fix the Hilbert space dimension of the environment to $d_B=d^n$, which one may interpret as $n$ interacting particles with local dimension $d$. In a celebrated work by Reebs and Wolf~\cite{reeb2014improved}, a finite-size correction for thermodynamic processes that reduce system's entropy, $\Delta S_{\ms{S}} \leq 0$ was derived, namely 
\begin{align}   
   \label{eq:rw_bound}
   \Sigma  \geq \frac{2 (\Delta S_{\ms{S}})^2}{\log^2 (d-1)+4} = \mathcal{O}\left(\frac{1}{n^2}\right). 
\end{align}
The above bound is valid for sufficiently large $n$ and applies to any environment, regardless of the specific system-environment interaction and process being implemented.

However, while Eq. \eqref{eq:rw_bound} puts a fundamental bound on the entropy production in \emph{any} thermodynamic process, it is not clear whether it can actually be achieved. More specifically,  it remains an open question whether the quadratic scaling from Eq. \eqref{eq:rw_bound} can be obtained with an actual thermodynamic process, or in fact any scaling beyond $\mathcal{O}(1/n)$.  
Our next result shows that this quadratic scaling  can be indeed achieved in a thermodynamic process. For that we will present an explicit process that realizes Landauer erasure.

Here a two-level quantum system prepared in the maximally mixed state $\rho_{\ms{S}} = \mathbb{1}_{\ms{S}}/2$  is mapped into a state close to the ground state. For that we use a thermodynamic process $\mathcal{P} = (\rho_{\ms{S}}, U, H_{\ms{B}})$ that produces the state
\begin{align}
    \sigma_{\ms{SB}} = U(\rho_{\ms{S}} \ot \tau_{\ms{B}})U^{\dagger},
\end{align}
where we want $\sigma_{\ms{S}} = (1-q) \dyad{0}_{\ms{S}} + q \dyad{1}_{\ms{S}}$ with $q\approx 0$. For the environment $\ms{B}$ we choose a system composed of $n$ qubits so that $d_{\ms{B}} = 2^n$.  Our main goal is to find a Hamiltonian $H_{\ms{B}}$ and a unitary interaction $U$ which implements the transition $\rho_{\ms{S}} \rightarrow \sigma_{\ms{S}}$ with entropy production as close as possible to the fundamental bound of Eq. \eqref{eq:rw_bound}. 

The unitary $U$ is chosen to be the so-called \emph{max-cooling unitary} \cite{clivaz2019unifying}. Such unitary reorders the eigenvalues $\{\lambda_{ij}\}$ of $\rho_{\ms{S}} \ot \tau_{\ms{B}}$, so that the largest of them is mapped to $\ket{0,0}_{\ms{SB}}$,  the second largest to $\ket{0,1}_{\ms{SB}}$ and so on until the whole subspace corresponding to the ground state $\ket{0}_{\ms{S}}$ is filled up with the largest eigenvalues. This operations reorders the set of eigenvalues into $\{\lambda_{ij}^{\downarrow}\}$, so that the final state of the system and environment after the action of $U$ can be written as
\begin{align} \label{eq:max_cool}
    \sigma_{\ms{SB}} = \sum_{\substack{i \in \{0,1\} \\ j \in \{0, 1, \ldots d_{\ms{B}-1}\} }} \lambda_{ij}^{\downarrow} \dyad{i,j}_{\ms{SB}}.
\end{align}
The max-cooling unitary maximizes the ground-state occupation of the system $\ms{S}$. In general we have no guarantee that the above unitary will be optimal for minimizing entropy production~\cite{Ralph2024}. Since we are interested in small $q$, we  make the cooling process explicitly depend on $n$ by taking $q = \frac{1}{2} n^{-\alpha}$ for some integer $\alpha$. 

The Hamiltonian of the environment is taken to be
\begin{align} \label{eq:inter_spec}
    H_{\ms{B}} =  \sum_{j=0}^{d_{\ms{B}}-1} E_{\ms{B}}^{(j)} \dyad{j}_{\ms{B}} = \sum_{i=0}^{n} \sum_{k=0}^{\Omega_{i}} \epsilon_i \dyad{\epsilon_i, k}. 
\end{align}
In order to find the degeneracy $\{\Omega_i\}$ and energy spectrum $\{\epsilon\}$ we perform numerical optimization using simulated annealing~\cite{kirkpatrick1983optimization}. Specifically, for small values of $n$ we observe that the optimal degeneracy can be well approximated by $\Omega_i = 2^{i-1}$ for $i \geq 1$ with $\Omega_0 = 1$. Furthermore, we find that the optimal energies $\{\epsilon_i\}$ can be described by
\begin{align}\label{eq:inter_e}
    \beta \epsilon_i = \log \Omega_{i+1} - \log \left[1 + n^{-\alpha} - \cos\left(\frac{2 \pi i}{n}\right)\right],
\end{align}
where $\alpha > 2$ is a real parameter that quantifies how close the final state is to the ground state. Indeed, the process $\mathcal{P}$ as defined above leads to the final state $\sigma_{\ms{S}}$ with $q = \frac{1}{2} n^{-\alpha}$. 

Given the above ansatz, in Appendix \ref{app2} we calculate the entropy production associated with the process $\mathcal{T}$  for $\alpha > 2$. We find that it can be bounded as
\begin{align} \label{eq:diss_protocol}
    \Sigma \geq 2 \pi^2 \frac{1}{n^2} + {\mathcal{O}}(1/n^2),
\end{align}
where $\mathcal{O}\left(1/n^2\right)$ indicates terms that vanish quicker than $n^{-2}$. We therefore see that entropy production decays \emph{quadratically} with the size of the environment. This is in stark contrast to the case of non-interacting environments [see Eq. \eqref{eq:noninter_bound}], where entropy production can decrease, at most, linearly with the size of the environment. This is the second main result of our paper, showcasing the potential role of interactions in decreasing entropy production in finite-size environments. 


\section{Presence of a thermal phase transition }

\begin{figure}
    \centering
    \includegraphics[width=\linewidth]{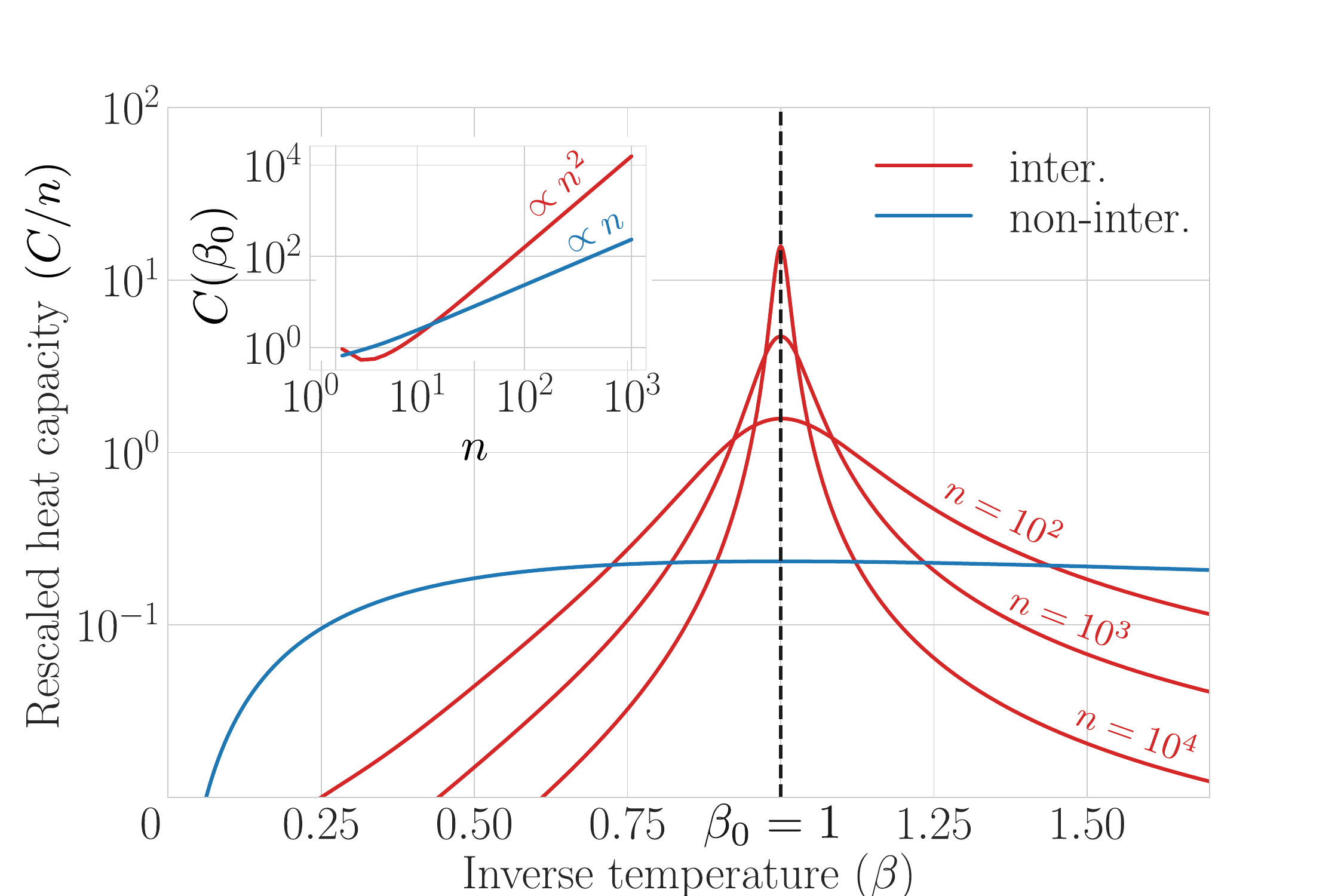}
    \caption{\textbf{Heat capacity close to a phase transition.} The main plot shows the rescaled heat capacity $C/n$ as a function of the inverse temperature for the Hamiltonian given by Eq. \eqref{eq:inter_spec} with parameter $\beta = \beta_0 = 1$. We observe that for the interacting environment (red curves) phase transition occurs precisely at $\beta = \beta_0$, whereas for the non-interacting environment the rescaled heat capacity $C/n$ approaches a constant value. The inset shows that $C$ diverges as $C \propto n^2$ as a function of the number of qubits $n$, whereas $C \propto n$ for a non-interacting environment.}
    \label{fig:2}
\end{figure}

Our previous results show, by explicitly constructing the environment's Hamiltonian and corresponding process, how to achieve a quadratic convergence to Landauer's bound in the environment's size. However, they provide little physical intuition on the origin of such advantages. To bring some insight into this question, let us connect our result to the presence of a thermal phase transition. For that, as in~\eqref{eq:nonint_firststep},  we define a density operator $\omega_{\ms{B}}$ such that $\omega_{\ms{B}} \propto e^{-\beta^{\star} H_{\ms{B}}}$ and $\beta^{\star}$ is chosen to satisfy  $\tr[H_{\ms{B}}\omega_{\ms{B}}] = \tr[H_{\ms{B}}\sigma_{\ms{B}}]$. Then, the following bound holds for any thermodynamic process $\mathcal{P} = (\rho_{\ms{S}}, U, H_{\ms{B}})$~\cite{reeb2014improved}, 
\begin{align}
\label{eq:heat_capacity}
\Sigma \geq \frac{(\beta Q)^2}{2 \max_{\gamma \in [\beta, \beta^*]}\mathcal{C}(\gamma) }
\end{align}
where $\mathcal{C}(\gamma) := \gamma^2 \langle (H_{\ms{B}} - \langle H_{\ms{B}}\rangle_{\gamma})^2\rangle_{\gamma}$ is the heat capacity of the thermal environment, and the average $\langle \cdot \rangle_{\gamma}$ is performed with respect to a Gibbs state at an inverse temperature $\gamma$. From this expression, it becomes clear that a faster convergence than $\mathcal{O}(1/n)$ is only possible for finite-size environments close to a critical point  in the sense of finite-size scaling theory~\cite{Suzuki1977,Fisher1972Scaling}, for which $\mathcal{C} \propto n^{1+x}$ with $x$ a critical exponent. In Fig.~\ref{fig:2} we show the heat capacity $\mathcal{C}$ of the model~\eqref{eq:inter_spec}, with energy spectrum given by Eq. \eqref{eq:inter_e} and parameter $\beta$ set to $\beta_0$. As expected, $\mathcal{C}$ diverges as $C\propto n^2$ precisely at~$\beta=\beta_0$, showcasing the presence of a thermal phase transition.  

From  the bound \eqref{eq:heat_capacity}, it naturally follows  that only environments at the verge of  a critical point can approach Landauer's bound faster than $\mathcal{O}(1/n)$. 
Note, however, that in principle non-critical interacting systems could  have a better prefactor than that from the lower bound   \eqref{eq:noninter_bound} for non-interacting systems. It is also important to realise that a phase transition is not sufficient for observing a super-extensive decay of entropy production. In particular, naively one could expect that the spectrum that maximises  $\mathcal{C}$~\cite{Correa2015Indivudal,Abiuso2024Optimal} is also optimal for cooling; this turns out to be false and in fact it does not even enable approaching Landauer's bound (see Appendix \ref{app:phase_transition_notsuff} for details). 

\section{Conclusion and outlook}

In this article, we considered the optimisation of thermodynamic processes, particularly Landauer erasure, in the presence of a finite-size environment. Our main goal was to explore the  fundamental limits that the environment's size imposes on thermodynamic reversibility.  First, we considered an environment made up of $n$ non-interacting particles, as commonly considered in the literature. We derived a fundamental bound that shows that the  entropy production $\Sigma$ can decay at most as $\mathcal{O}(1/n)$, see Eq.~\eqref{eq:noninter_bound}. Instead, we showed that interacting environments can surpass this bound, and derived an explicit process for which $\Sigma \propto \mathcal{O}(1/n^2)$, see Eq. ~\eqref{eq:diss_protocol}. This is in fact the best possible scaling of the entropy production with $n$, which follows from the lower bound  derived in Ref.~\cite{reeb2014improved}--see also Eq.~\eqref{eq:rw_bound}. 
Our work hence provides new insights on the potential of interactions and criticality for minimizing entropy production and increasing thermodynamic efficiency~\cite{Allahverdyan2013Carnot,campisi2016power,Abiuso2020,Barra2022,Rolandi2023Collective,Namede2023Thermodynamics,Buffoni2023}.

While the protocol we derived requires highly engineered Hamiltonians and interactions, our results suggest the possibility of decreasing such demands. Indeed, we showed that any such advantage requires the reservoir to be close to a thermal phase transition, so that its heat capacity diverges as $n^{1+x}$ with $x>0$. In this sense, it would be interesting to explore the potential of locally interacting reservoirs, such as Ising models close to criticality. On the other hand, we also noted that criticality is a necessary but no sufficient condition for a faster convergence to the reversible limit. It would be hence interesting to identify the necessary and sufficient requirements of the reservoir for such advantages.

There are also several interesting connections between this work and recent literature, which represent exciting directions to explore in the future. A first one is connecting these finite-size effects to finite-time (quantum) stochastic thermodynamics. For collisional models~\cite{reeb2014improved,skrzypczyk2014work,Baumer2019imperfect}, time $t$ and the bath's size are linearly related, leading to a decay of the entropy production as~$\Sigma \propto 1/t$. This scaling is in fact naturally found in the so-called finite-time Landauer principle, analyzed both in classical and quantum setups~\cite{Proesmans2020,VanVu2022,Lee2022,Ma2022,Zhen2021,rolandi2022finite,vanvu2023}. Our result then naturally opens the question whether 
  environments at the verge of a phase transition can lead to a faster decay of the entropy production, even leading to $\Sigma \propto 1/t^2$. 

A second exciting avenue is the connection of these results with complexity. Recent results suggest that the standard time-energy tradeoff between resources in thermodynamics can be extended with a third resource, namely the complexity of the allowed operations~\cite{Taranto2023Landauer}. Arguably, achieving the scaling from Eq.~\eqref{eq:diss_protocol} requires not only preparing the bath in an  engineered interacting Hamiltonian, but also highly collective unitary operations. These operations can be implemented with many-body interactions or with more realistic two-body interactions but at the cost of a higher time~\cite{nielsen2010quantum}. 
Investigating such questions  could provide new insights on the role of complexity in thermodynamics. 

\begin{acknowledgments}
We thank Ralph Silva and Pharnam Bakhshinezhad for stimulating discussions. 
P. L.-B. and M. P.-L. acknowledge the Swiss National Science Foundation for financial support through NCCR SwissMAP and Ambizione grant PZ00P2-186067, respectively. M. P.-L also acknowledges funding from the Spanish Agencia Estatal de Investigacion through the grant  ``Ram{\'o}n y Cajal RYC2022-036958-I''.
\end{acknowledgments}

\bibliographystyle{apsrev4-2}
\bibliography{citations}

\appendix
\widetext

\section{Entropy production in non-interacting thermal environments} \label{app1}
Consider a system $\ms{S}$ prepared in a quantum state $\rho_{\ms{S}}$. Let this system interact with a thermal environment composed of non-interacting particles each described by a single-particle Hamiltonian $h_i$ for $i = 1, \ldots, n$. The total Hamiltonian of the environment is therefore $H_{\ms{B}} = \sum_{i=1}^n h^{(i)}_{\ms{B}}$. Consequently, the thermal state of the environment can be written as $\tau_{\ms{B}} = \bigotimes_{i=1}^n \tau_{\ms{B}}^{(i)}$. The systems $\ms{S}$ and $\ms{B}$ interact via a unitary $U$ so that
\begin{align} \label{eq:app_A0}
    \rho_{\ms{S}} \ot \tau_{\ms{B}} \rightarrow \sigma_{\ms{SB}} := U[\rho_{\ms{S}} \ot \tau_{\ms{B}}]U^{\dagger}.
\end{align}
The entropy production $\Sigma$ in any such unitary process is given by
\begin{align}
    \Sigma = I(\ms{S}:\ms{B})_{\sigma} + D(\sigma_{\ms{B}} \|  \tau_{\ms{B}}).
\end{align}
Let us now introduce another Gibbs state of the environment $\ms{B}$ (with a potentially different temperature) with energy equal to $\sigma_{\ms{B}}$, that is
\begin{align}
    \omega_{\ms{B}} := \frac{1}{Z} e^{-\beta' H_{\ms{B}}} \quad \text{with} \quad Z' = \tr[e^{-\beta' H_{\ms{B}}}] \quad \text{such that} \quad \tr[\omega_{\ms{B}}H_{\ms{B}}] = \tr[\sigma_{\ms{B}}H_{\ms{B}}]. 
\end{align}
Since the Gibbs state is, by definition, a state with maximal von Neumann entropy given fixed average energy, we have $S(\omega_{\ms{B}}) \geq S(\sigma_{\ms{B}})$. Due to the additivity of the von Neuman entropy and unitarity of the thermodynamic process \eqref{eq:app_A0} we also have $\Delta S_{\ms{S}} + \Delta S_{\ms{B}} \geq 0$, which further implies that
\begin{align} \label{eq:app_A5}
     S(\omega_{\ms{B}}) - S(\tau_{\ms{B}}) \geq S(\sigma_{\ms{B}}) - S(\tau_{\ms{B}}) \geq - \Delta S_{\ms{S}}
\end{align}

Importantly, the relative entropy obeys the following identity \cite{csiszar1975}
\begin{align}
    D(\sigma_{\ms{B}}\| \tau_{\ms{B}}) =  D(\sigma_{\ms{B}}\| \omega_{\ms{B}}) +  D(\omega_{\ms{B}}\| \tau_{\ms{B}}) \geq  D(\omega_{\ms{B}}\| \tau_{\ms{B}}),
\end{align}
which, given that both $I(\ms{S}:\ms{B})_{\sigma}$ and $D(\sigma_{\ms{B}}\|\omega_{\ms{B}})$ are non-negative, leads to the lower bound $\Sigma \geq D(\omega_{\ms{B}}\| \tau_{\ms{B}})$. Using the fact that the environment $\ms{B}$ is composed of $n$ non-interacting particles we can further write 
\begin{align} \label{eq:app_diss1}
    \Sigma \geq \sum_{i=1}^n D(\omega_{\ms{B}}^{(i)}\| \tau_{\ms{B}}^{(i)}).
\end{align}
Now we recall a general lower bound for the relative entropy introduced in Ref. \cite{reeb2015tight}. For that we use an auxiliary function $M(x, y)$ which for $y \geq 2$ and $x \in [-\log y, \log y]$  is defined as
\begin{align}
    M(x, y) := \min_{0 \leq a,b \leq (d-1)/d} \left\{ s(a,b) \,|\, h(a) - h(b) 
+ (a-b) \log (d-1) = x\right\},
\end{align}
where $h(a) := - a \log a - (1-a) \log (1-a)$ and $s(a,b) := a (\log a - \log b) + (1-a) [\log(1-a) - \log(1-b)]$. 

The function $M(x,y)$ provides a tight lower bound for the relative entropy. More specifically, for any two density operators $\rho$ and $\sigma$ of dimension $d$ with $2 \leq d < \infty$ and $\Delta := S(\rho) - S(\sigma)$ we have \cite{reeb2015tight}:
\begin{align} \label{eq:app_A1}
    D(\rho \| \sigma) \geq M(\Delta,d).
\end{align}

Importantly, Ref. \cite{reeb2015tight} showed that $M(x,y)$ satisfies the bound 
\begin{align}
\label{eq:app_A2}
M(x,y) \geq x^2 / (3 \log^2 y).    
\end{align}
Moreover, the quantity $M(x,y)$ is convex in its first argument, that is for $\bar{x} = p x_1 + (1-p) x_2$ with $0 \leq p \leq 1$ it satisfies $M(\bar{x}, y) \leq p M(x_1, y) + (1-p) M(x_2, y)$. This last property in particular implies that
\begin{align}
    \label{eq:app_A3}
    \sum_{i=1}^n M(x_i, y) \geq n M\left(\frac{1}{n}\sum_{i=1}^n x_i, y\right).
\end{align}

Let us now return to our main problem from this section, i.e. lower-bounding the entropy production from Eq. \eqref{eq:app_diss1}. Introducing $\Delta_i := S(\omega_{\ms{B}}^{(i)}) - S(\tau^{(i)}_{\ms{B}})$ and $\Delta S_{\ms{B}} := \sum_{i=1}^n \Delta_i$ we can write
\begin{align}
    \Sigma &\geq \sum_{i=1}^n D(\omega_{\ms{B}}^{(i)}\|\tau_{\ms{B}}^{(i)}) \\
    &\geq \sum_{i=1}^n M(\Delta_i, d) \\
    &\geq n M\left(\frac{1}{n} \sum_{i=1}^n \Delta_i, d\right) \\
    & \geq n \left(\frac{\Delta S_{\ms{B}}}{n}\right)^2 \frac{1}{3 \log^2 d} \\
    &= \frac{1}{n} \frac{\Delta S_{\ms{B}}^2}{3 \log^2 d}.
\end{align}
Where in the second line we used Eq. \eqref{eq:app_A1}, in the third line we used Eq. \eqref{eq:app_A3} and in the fourth line we used Eq. \eqref{eq:app_A2}. Finally, using Eq. \eqref{eq:app_A5} we may conclude that 
\begin{align}
    \Sigma \geq \frac{1}{n}  \frac{\Delta S_{\ms{S}}^2}{3\log^2 d}.
\end{align}


\section{Landauer erasure in interacting thermal environments} \label{app2}
Consider a system $\ms{S}$ with Hamiltonian $H_{\ms{S}}$ to be prepared in a quantum state $\rho_{\ms{S}}$. Let this system interact with an environment in a state $\tau_{\ms{B}}(H_{\ms{B}})$ via a unitary $U$, that is
\begin{align} \label{eq:app1}
    \rho_{\ms{S}} \ot \tau_{\ms{B}}(H_{\ms{B}}) \rightarrow \sigma_{\ms{SB}} := U[\rho_{\ms{S}} \ot \tau_{\ms{B}}(H_{\ms{B}})]U^{\dagger}.
\end{align}
We can always rotate the system $\ms{S}$ into the energy eigenbasis, i.e. the basis specified by ${H}_{\ms{S}}$. Therefore it is enough to consider the arising energy probability distribution $\bm{p}_{\ms{S}} := (p_1, p_2, \ldots, p_n)$, where $\{p_i\}_{i=1}^{d}$ are the eigenvalues of $\rho_{\ms{S}}$. Without loss of generality we can further assume that the unitary is a permutation of eigenvalues. Therefore the problem of transforming a quantum system from Eq. \eqref{eq:app1} can be without loss of generality written as
\begin{align}
    \bm{p}_{\ms{S}} \ot \bm{g}_{\ms{B}}(H_{\ms{B}}) \rightarrow \Pi [\bm{p}_{\ms{S}} \ot \bm{g}_{\ms{B}}(H_{\ms{B}})],
\end{align}
where $\bm{g}_{\ms{B}}(H_{\ms{B}}) := \text{diag}[\tau_{\ms{B}}(H_{\ms{B}})]$ is a probability vector formed from the eigenvalues of the Gibbs state $\tau_{\ms{B}}(H_{\ms{B}})$. For the environment $\ms{B}$ we choose a system composed of $D = 2^n$ energy levels, i.e.
\begin{align}
    H_{\ms{B}} = \sum_{i=0}^{n} \sum_{k=0}^{\Omega_{i}} \epsilon_i \dyad{\epsilon_i, k}, \qquad \Omega_i = \begin{cases}
        1 \quad \text{for} \quad i = 0,\\
        2^{i-1} \quad \text{for} \quad i \geq 1.
    \end{cases}
\end{align}
We further assume that $\epsilon_0 = 0$ without loss of generality. The permutation $\Pi$ is chosen so that to maximize the ground state occupation of $\ms{S}$. In other words, its action is to sort the vector $\bm{p}_{\ms{S}} \ot \bm{g}_{\ms{B}}(H_{\ms{B}})$ in such a way that $\langle 0| \Pi[\bm{p}_{\ms{S}} \ot \bm{g}_{\ms{B}}(H_{\ms{B}})]|0 \rangle$ is maximized, i.e.
\begin{align}
\Pi = U = \text{max-cooling permutation (to discuss)}.    
\end{align}
Let us define $g_i := e^{-\beta \epsilon_i}/Z_n$, where $Z_n := \sum_{i=0}^n \Omega_i e^{-\beta \epsilon_i}$. Visually the action of the permutation $\Pi$ on the initial state of the system and the environment can be depicted as in Fig. (?).

The energy change on the environment $\ms{B}$ as a result of applying the transformation from Eq. (\ref{eq:app1}). It is given by $\Delta E_{\ms{B}} = E_{\ms{B}}^{(f)} - E_{\ms{B}}^{(i)}$, where the initial and final energies are respectively given by
\begin{align} \label{eq:app_energies}
    E_{\ms{B}}^{(i)} &= \sum_{i=0}^n \Omega_i g_i \epsilon_i, \qquad E_{\ms{B}}^{(f)} = \frac{1}{2} \Omega_0 g_0 \epsilon_0 +\frac{1}{2} \sum_{i=1}^n \Omega_i g_{i-1} \epsilon_i + \frac{1}{2} \sum_{i=0}^n \Omega_i g_n \epsilon_i,
\end{align}
Consequently, the total energy change of the environment under the action of $\Pi$ can be expressed as
\begin{align}\label{eq:app_dE}
     \Delta E_{\ms{B}} &= \frac{1}{2} \sum_{i=1}^n \Omega_i (g_{i-1} - 2g_i+g_n)\epsilon_i + \frac{1}{2}\Omega_0 (g_n-g_0)\epsilon_0.
\end{align}
Due to the evolution specified by $\Pi$, the system $\ms{S}$ transforms as
\begin{align}
    \bm{p}_{\ms{S}} \rightarrow \Pi[\bm{p}_{\ms{S}} \ot \bm{g}_{\ms{B}}(H_{\ms{B}})] = (1-\Omega_{n}g_n, \Omega_n g_n).
\end{align}
Let us now rewrite Eq. \eqref{eq:app_energies} using a more convenient parametrization for energies $\{\epsilon_i\}$, that is:
\begin{align}
    \beta \epsilon_i = \log \Omega_{i+1} - \log r_i,
\end{align}
where $r_i > 0$ for all $i \in \{1, \ldots, n\}$ are arbitrary real numbers and $\beta \epsilon_0 := - \log r_0$. Note that the above parametrization is without loss of generality. Consequently, we can rewrite Eq. \eqref{eq:app_dE} as
\begin{align}   \label{eq:app_de2} \nonumber
    \beta \Delta E_{\ms{B}} &= \frac{1}{2Z} \sum_{i=1}^n (r_{i-1} - r_i + \Omega_{i-n}r_n) ( \log \Omega_{i+1} - \log r_i) - \frac{1}{2Z}  \left(2^{-n} r_n - r_0\right) \log r_0 \\
    &= \underbrace{\frac{1}{2Z} \sum_{i=1}^n (r_{i-1} - r_{i} + \Omega_{i-n} r_n) \log \Omega_{i+1}}_{A} - \underbrace{\frac{1}{2Z}\sum_{i=1}^n (r_{i-1}-r_i) \log r_i}_{B} -  \underbrace{\frac{1}{2Z}\sum_{i=1}^n \Omega_{i-n}r_n \log r_i}_C  \\ &\quad - \underbrace{\frac{1}{2Z}  \left(2^{-n} r_n - r_0\right) \log r_0}_{D}.
     \nonumber
\end{align}

Let us now choose a specific set of parameters $r_i$, namely
\begin{align} \label{eq:app_rdef}
    r_i = 1 + n^{-\alpha} - \cos\left(\frac{2 \pi i}{n}\right),
\end{align}
where $\alpha \geq 0$ is a real parameter. Notice that $r_0 = r_n = n^{-\alpha}$. Moreover, we can now also compute the partition function $Z$, namely
\begin{align}
Z = \sum_{i=0}^n \Omega_i e^{-\beta \epsilon_i} = r_0 + \frac{1}{2}\sum_{i=1}^n  r_i = n^{-\alpha} +\frac{1}{2} n(1+n^{-\alpha}) = \frac{1}{2}\left(n+n^{1-\alpha} + 2n^{-\alpha}\right) = \mathcal{O}\left(n\right).
\end{align}
where we used the fact that $\sum_{i=1}^n \cos\left(\frac{2\pi i}{n}\right) = 0$. With this we can now upper bound the different sums appearing in Eq. \eqref{eq:app_de2}. Specifically, let us start with 
\begin{align}
    A &= \frac{1}{2Z} \sum_{i=1}^n (r_{i-1} - r_{i} + \Omega_{i-n} r_n) \log \Omega_{i+1} \\
    &= \frac{1}{2Z} \sum_{i=1}^n i \left[\cos\left(\frac{2\pi i}{n}\right) - \cos\left(\frac{2\pi (i-1)}{n}\right) + 2^{i-n-1}n^{-\alpha}\right] \log 2 \\
    &= \frac{1}{2 Z}\left(n+n^{-\alpha} (2^{-n} + n - 1) \right) \log 2 \\
    &= \left(1 - \frac{3}{n^{\alpha+1}+n+2}\right) \log 2 \\
    &\leq \log 2,
\end{align}
where we used the closed forms of the following summations 
\begin{align}
    \sum_{i=1}^n i \cos\left(\frac{2\pi i}{n}\right) = - \sum_{i=1}^n i \cos\left(\frac{2\pi (i-1)}{n}\right) = \frac{1}{2}n \qquad \text{and} \qquad \sum_{i=1}^n i\, 2^{i} = 2 \left[1 + 2^n(n-1)\right].
\end{align}
Let us now calculate the second labelled term from Eq. \eqref{eq:app_de2}. For that we expand $r_{i-1}$ in the powers of $1/n$, namely 
\begin{align}
    r_{i-1} = r_i - 2 \pi \sin\left(\frac{2 \pi i}{n}\right) \frac{1}{n} + 2 \pi^2 \cos\left(\frac{2 \pi i}{n}\right) \frac{1}{n^2} + \On{3}.
\end{align}
Consequently we can write the second term from Eq. \eqref{eq:app_de2} as
\begin{align}
    B &= \frac{1}{2Z}\sum_{i=1}^n (r_{i-1}-r_i) \log r_i \\ 
    &=  \frac{1}{2Z} \frac{2 \pi^2}{n^2} \sum_{i=1}^n \cos \left(\frac{2 \pi i}{n}\right) \log \left(1 + n^{-\alpha} - \cos \left(\frac{2 \pi i}{n}\right)\right) + \On{3} \\
    & = \frac{1}{Z} \frac{\pi^2}{n^2} \sum_{i=0}^n \cos \left(\frac{2 \pi i}{n}\right) \log \left(1 + n^{-\alpha} - \cos \left(\frac{2 \pi i}{n}\right)\right) + \frac{\alpha}{2Z}\frac{2 \pi^2}{n^2}\log\left(n\right) + \On{3} \\
    &\geq \frac{1}{Z} \frac{\pi^2}{n^2} \int_{0}^n \cos \left(\frac{2 \pi x}{n}\right) \log \left(1 + n^{-\alpha} - \cos \left(\frac{2 \pi x}{n}\right)\right) \text{d}x + \mathcal{O}\left(\frac{\log n}{n^3}\right). \label{eq:app_B4}
\end{align}
In the second line we used the fact that $\sum_{i=1}^n \sin \left(\frac{2 \pi i}{n}\right) \log r_i =0$ as it is a sum of a product of (shifted) even and odd functions. In the fourth line we used the following property of the Riemann integral:
\begin{align}
    \sum_{i=k+1}^n f(i) \leq \int_{k}^n f(x) \text{d}x \leq \sum_{i=k}^n f(i).
\end{align}
The integral appearing in Eq. \eqref{eq:app_B4} can be computed analytically by observing that
\begin{align}
    F(x, a, b) &:= \int \cos \left(a x\right) \log \left[b - \cos \left(a x\right)\right] \text{d}x \\
    &= \frac{2 \sqrt{1-b^2}}{a} \tanh^{-1}\left(\frac{(1+b) \tan \left(\frac{a x}{2}\right)}{\sqrt{1-b^2}}\right)+\sin\left(a x\right) \left[\log (b-\cos (a x))-1\right] - b x.
\end{align}
Specifically, using the above in Eq. \eqref{eq:app_B4} leads to 
\begin{align}
    B &\geq \frac{1}{Z} \frac{\pi^2}{n^2} \left[F\left(n, \frac{2 \pi}{n}, 1 + n^{-\alpha}\right) - F\left(0, \frac{2 \pi}{n}, 1 + n^{-\alpha}\right)\right] + \mathcal{O}\left(\frac{\log n}{n^3}\right) \\
    &=  -\frac{1}{Z} \frac{\pi^2}{n^2} n (1 + n^{-\alpha}) + \mathcal{O}\left(\frac{\log n}{n^3}\right) \\
    &= -2\pi^2 \frac{1}{n^2 + \frac{2n}{1+n^{\alpha}}} + \mathcal{O}\left(\frac{\log n}{n^3}\right) \\
    &= -2 \pi^2 \frac{1}{n^2} + \mathcal{O}\left(\frac{\log n}{n^3}\right),
\end{align}
where in the last line we used the fact that $\left(n^2 + \frac{2n}{1+n^{\alpha}}\right)^{-1} = n^{-2} + \mathcal{O}(n^{-3})$ for $\alpha \geq 0$. 

Let us now consider the third term appearing in Eq. \eqref{eq:app_de2}, namely
\begin{align}
    C &= \frac{1}{2Z}\sum_{i=1}^n \Omega_{i-n}r_n \log r_i \\
    & \geq \frac{1}{2Z} r_n \log r_n \sum_{i=1}^n \Omega_{i-n} \\
    &= - \frac{1}{2Z} (1-2^{-n}) \alpha\, n^{-\alpha} \log n  \\
    &= \alpha \frac{\log n}{n^{1+\alpha}} + \mathcal{O}\left(\frac{\log n}{n^{2}} \frac{1}{n^{\alpha}}\right).
\end{align}
where we used the fact that $\log r_i \geq \log r_n $ for $i \in 0, 1, \ldots, n$.

Finally, the last term in Eq. \eqref{eq:app_de2} can be written as
\begin{align}
    D &= \frac{1}{2Z}  \left(2^{-n} r_n - r_0\right) \log r_0 \\
    &= \frac{\alpha \log n}{2 + n + n^{1+\alpha}} + \mathcal{O}\left(n^{-\alpha}e^{-n}\right)\\
    &= \alpha \frac{\log n}{n^{1+\alpha}} + \mathcal{O}\left(\frac{\log n}{n^{2}} \frac{1}{n^{\alpha}}\right).
\end{align}
Combining our bounds for the terms appearing in Eq. \eqref{eq:app_de2} we can write
\begin{align}
    \Delta E_{\ms{B}} &= A - B - C - D \\
    &\leq \log 2 + 2 \pi^2 \frac{1}{n^2} - 2 \alpha \frac{\log n}{n^{1+\alpha}} + \mathcal{O}\left(\frac{\log n}{n^2} \frac{1}{n^{\min(1, \alpha)}}\right).
\end{align}
Let us now label with $q$ the excited state occupation of the final state of the system. Observe that it can be written as
\begin{align}
    q := \Omega_n g_n = \frac{\Omega_n}{\Omega_{n+1}} r_n = \frac{1}{2} n^{-\alpha}.
\end{align}
This allows us to write the entropy change of the system $\Delta S$ as
\begin{align}
    \Delta S &:= -q \log q - (1-q) \log (1-q) - \log 2 \\
    &= n^{-\alpha} \arctan\left(1-n^{-\alpha}\right) - \log\left(1-\frac{1}{2} n^{-\alpha}\right) - \log 2 + \On{4} \\
    &= \frac{1}{2}\left(1 + \log 2\right) n^{-\alpha} + \frac{1}{2} \alpha n^{-\alpha} \log n -\log 2 + \mathcal{O}\left(\frac{1}{n^{\min(1+\alpha, 4)}}\right).
\end{align}
This allows us to upper bound the entropy production as
\begin{align}
    \Sigma &:= \Delta E_{\ms{B}} + \Delta S \\
    &\leq  2\pi^2 \frac{1}{n^2} + \frac{1}{2} \frac{1}{n^{\alpha}} (1+\log 2) + \alpha \frac{1}{2} \left(1-\frac{4}{n}\right)\frac{1}{n^{\alpha}} \log n + \mathcal{O}\left(\frac{1}{n^{\min(1+\alpha, 3)}}\right) \\
    & = \begin{cases}
        2\pi^2 \frac{1}{n^2} + \mathcal{O}\left(\frac{\log n}{n^{\alpha}}\right)  &\text{for} \quad\alpha  \leq 3, \\
        2\pi^2 \frac{1}{n^2} + \mathcal{O}\left(\frac{1}{n^{3}}\right) &\text{for}\quad \alpha > 3.
    \end{cases} 
\end{align}
Observe further that by taking $\alpha > 2$ we have
\begin{align}
    \Sigma &\leq  2\pi^2 \frac{1}{n^2} + \mathcal{O}\left(1/n^2\right),
\end{align}
where the notation $\mathcal{O}\left(1/n^2\right)$ indicates terms that vanish quicker than $n^{-2}$, i.e. satisfy $\lim_{n\rightarrow\infty} [n^2 \cdot \mathcal{O}\left(1/n^2\right)] = 0$.

\section{Criticality is not sufficient for quadratic scaling of entropy production}
\label{app:phase_transition_notsuff}
In this Appendix we prove that being close to phase transition is not sufficient to observe a quadratic scaling of entropy production in thermodynamic processes. For that we will consider the task of Landauer erasure realized using a thermal environment at the verge of phase transition, and compare it to the process described in the main text (see also Appendix \ref{app2}). For that, we consider the spectrum that maximises the heat capacity~\cite{Correa2015Indivudal,Abiuso2024Optimal}, and investigate how it performs in terms of cooling a qubit. 

Consider a thermal environment at inverse temperature $\beta$ composed of $n$ qubits described by Hamiltonian $H_{\ms{B}}^{\text{deg}}$ such that
\begin{align}
    H_{\ms{B}} = \epsilon_0 \dyad{0}_{\ms{B}} + \epsilon \sum_{i=1}^{N} \dyad{i}_{\ms{B}},
\end{align}
where $N = 2^n - 1$. In what follows, without loss of generality, we assume $\epsilon_0 = 0$. As shown in Ref.~\cite{Correa2015Indivudal}, this Hamiltonian allows for a thermal phase transition which is manifested by a quadratic scaling of heat capacity, namely
\begin{align} \label{eq:quadratic_heat_capacity}
    \mathcal{C} = \frac{1}{4} n^2 \log^2 d,
\end{align}
which is in contrast with the typical extensive behaviour of the heat capacity (i.e. linear in $n$) for non-interacting environments. 

Suppose we couple unitarily the thermal environment with a qubit prepared in the maximally-mixed state, namely
\begin{align}
    \sigma_{\ms{SB}} = U\left(\frac{\mathbb{1}_{\ms{S}}}{2} \ot \gamma_{\ms{B}}\right)U^{\dagger},
\end{align}
where $\gamma_{\ms{B}} = \text{exp}(-\beta H_{\ms{B}})/\Tr[\text{exp}(-\beta H_{\ms{B}})]$. Our goal, just as before, is to achieve minimal entropy production for a given ground state occupation on $\ms{S}$. 

Let us denote the diagonal of $\gamma_{\ms{B}}$ with $\bm{g} := [a, (1-a)/N, (1-a)/N, \ldots (1-a)/N]$ and the diagonal of the initial state of the system as $\bm{p} = (1/2, 1/2)$. Consider permuting the elements of $\bm{p} \ot \bm{g}$. It can be shown by direct calculation that for $a \geq 2^{-n}$ the permutation which achieves minimal entropy production $\Sigma$ given fixed target state on the system achieves
\begin{align}
    \Tr_1[\Pi(\bm{p}\ot\bm{g})] &= \left(\frac{1-a}{2N} + \frac{1}{2}a, \frac{1-a}{2N} + \frac{1}{2}a, \frac{1-a}{N}, \ldots, \frac{1-a}{N}\right) := \bm{g}', \\
    \Tr_2[\Pi(\bm{p}\ot\bm{g})] &= (q, 1-q) =: \bm{q},
\end{align}
where $q = (1-x)/2$ with $x := [a - (1-a)/N]$ and $\epsilon = \epsilon(\beta_0) = [\log N - \log(\frac{1}{a}-1)]/\beta_0$. 
Furthermore we have $\beta Q = \epsilon \gamma/2$, while that the entropy production $\Sigma$ is given by Eq. \eqref{eq:entr_prod}, namely $\Sigma = \beta Q + \Delta S$.

\begin{figure}
    \centering
    \includegraphics[width=0.7\linewidth]{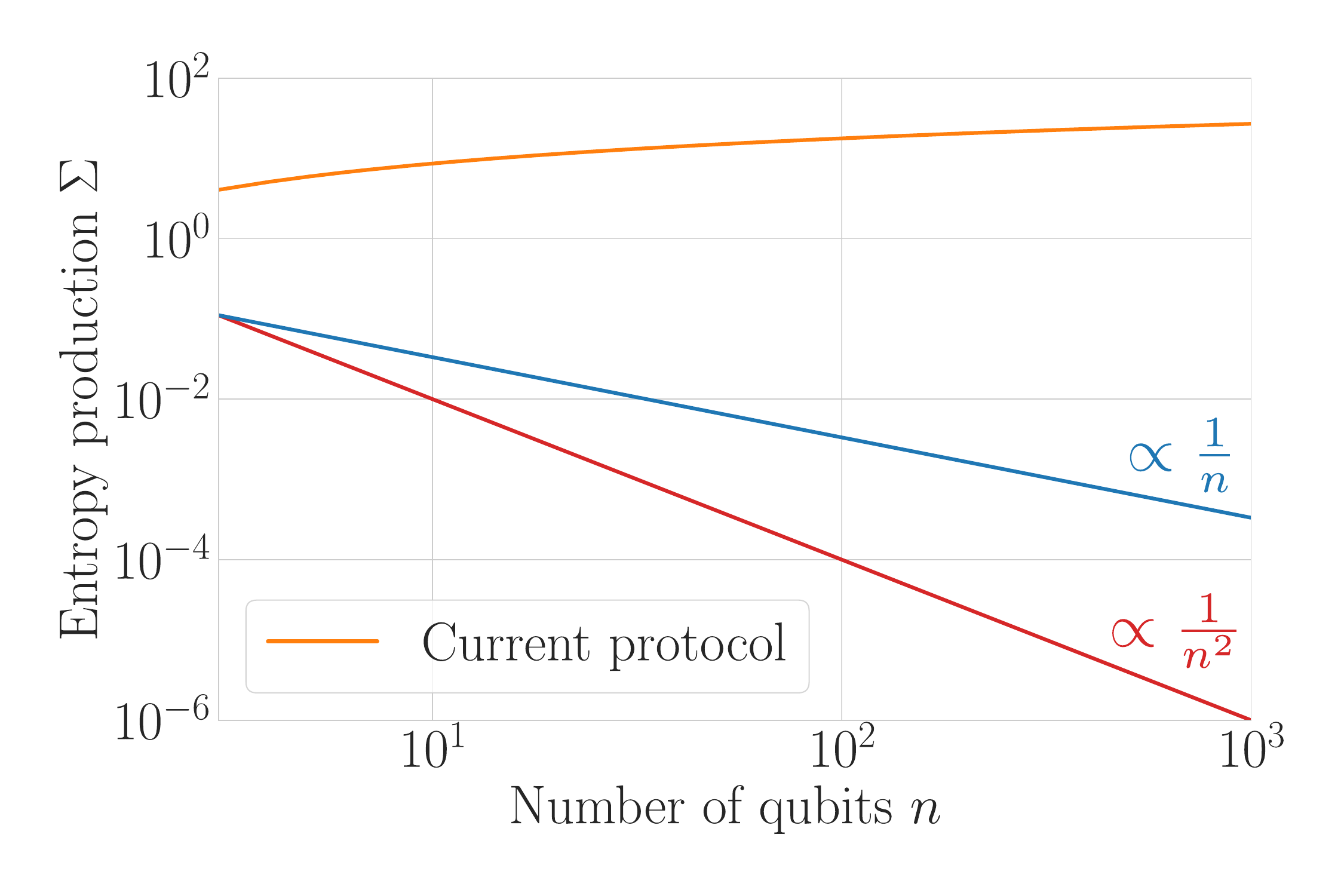}
    \caption{\textbf{Entropy production at criticality.} The panel shows the entropy production during Landauer erasure as a function of the number of qubits $n$, realised via the optimal protocol discussed in Appendix \ref{app:phase_transition_notsuff}. The environment used in the protocol clearly exhibits a thermal phase transition, as demonstrated by the super-extensive scaling of heat capacity [see Eq. \eqref{eq:quadratic_heat_capacity}]. For comparison we also plot the curves corresponding to the linear (blue) and quadratic (red) decay of entropy production. We see that the presence of a phase transition is not sufficient to observe a quadratic scaling of entropy production. }
    \label{fig:3}
\end{figure}

In Fig. \ref{fig:3} we compare the entropy production during Landauer erasure realised via the thermodynamic process discussed in this section. We also compare it with the quadratic scaling of entropy production observed in the protocol discussed in the main text. In particular, we observe that for the protocol discussed in this section the entropy production decreases linearly with $n$, namely $\Sigma \approx 1/n$. This demonstrates that criticality is not sufficient to experience a quadratic scaling of entropy production. 

\end{document}